\documentclass[journal]{IEEEtran}

\ifCLASSINFOpdf

\else

\fi

\usepackage{graphicx}
\usepackage[dvipsnames]{xcolor}
\usepackage{academicons} 
\newcommand{\orcid}[1]{\href{orcid.org/#1}{\textcolor[HTML]{A6CE39}{\aiOrcid}}}
\usepackage{scalerel}
\usepackage{tikz}
\usetikzlibrary{svg.path}

\definecolor{orcidlogocol}{HTML}{A6CE39}
\tikzset{
  orcidlogo/.pic={
    \fill[orcidlogocol] svg{M256,128c0,70.7-57.3,128-128,128C57.3,256,0,198.7,0,128C0,57.3,57.3,0,128,0C198.7,0,256,57.3,256,128z};
    \fill[white] svg{M86.3,186.2H70.9V79.1h15.4v48.4V186.2z}
                 svg{M108.9,79.1h41.6c39.6,0,57,28.3,57,53.6c0,27.5-21.5,53.6-56.8,53.6h-41.8V79.1z M124.3,172.4h24.5c34.9,0,42.9-26.5,42.9-39.7c0-21.5-13.7-39.7-43.7-39.7h-23.7V172.4z}
                 svg{M88.7,56.8c0,5.5-4.5,10.1-10.1,10.1c-5.6,0-10.1-4.6-10.1-10.1c0-5.6,4.5-10.1,10.1-10.1C84.2,46.7,88.7,51.3,88.7,56.8z};
  }
}

\newcommand\orcidicon[1]{\href{https://orcid.org/#1}{\mbox{\scalerel*{
\begin{tikzpicture}[yscale=-1,transform shape]
\pic{orcidlogo};
\end{tikzpicture}
}{|}}}}

\usepackage{multirow}
\usepackage{blindtext}
\usepackage{cite}
\usepackage{float}
\usepackage{stfloats}
\usepackage[colorlinks=true,linkcolor=black,anchorcolor=black,citecolor=black,filecolor=black,menucolor=black,runcolor=black,urlcolor=black]{hyperref}

\graphicspath{ {figures/} }

\begin{document}

\title{Investigating the Cognitive Response of Brake Lights in Initiating Braking Action Using EEG}

\author{Ramaswamy Palaniappan$^{\textsuperscript{\orcidicon{0000-0001-5296-8396}}}$\,~\IEEEmembership{Senior Member~IEEE}, Surej Mouli$^{\textsuperscript{\orcidicon{0000-0002-2876-3961}}}$\,~\IEEEmembership{Senior Member~IEEE}, Howard Bowman and Ian McLoughlin$^{\textsuperscript{\orcidicon{0000-0001-7111-2008}}}$\,~\IEEEmembership{Senior Member~IEEE} 

\thanks{R. Palaniappan is with the Data Science Research Group, School of Computing, University of Kent, UK (e-mail:\href{mailto:R.Palani@kent.ac.uk}{R.Palani@kent.ac.uk}).}

\thanks{S. Mouli is with the School of Engineering and Applied Science, Aston University, UK (e-mail:\href{mailto:surej@ieee.org}{s.mouli@aston.ac.uk}).}

\thanks{H. Bowman is with School of Computing, University of Kent and School of Psychology, University of Birmingham, UK (email:\href{mailto:H.Bowman@kent.ac.uk}{H.Bowman@kent.ac.uk}).}

\thanks{I. McLoughlin is with ICT Cluster, Singapore Institute of Technology, Singapore (email:\href{mailto:Ian.McLoughlin@singaporetech.edu.sg}{Ian.McLoughlin@singaporetech.edu.sg}).}

}

\maketitle

\begin{abstract}
Half of all road accidents result from either lack of driver attention or from maintaining insufficient separation between vehicles. Collision from the rear, in particular, has been identified as the most common class of accident in the UK, and its influencing factors have been widely studied for many years.
Rear-mounted stop lamps, illuminated when braking, are the primary mechanism to alert following drivers to the need to reduce speed or brake.
This paper develops a novel brain response approach to measuring subject reaction to different brake light designs. A variety of off-the-shelf brake light assemblies are tested in a physical simulated driving environment to assess the cognitive reaction times of $22$ subjects. 
Eight pairs of LED-based and two pairs of incandescent bulb-based brake light assemblies are used and electroencephalogram (EEG) data recorded. Channel Pz is utilised to extract the P3 component evoked during the decision making process that occurs in the brain when a participant decides to lift their foot from the accelerator and depress the brake. EEG analysis shows that both incandescent bulb-based lights are statistically slower to evoke cognitive responses than all tested LED-based lights. Between the LED designs, differences are evident, but not statistically significant, attributed to the significant amount of movement artifact in the EEG signal.

\end{abstract}


\IEEEpeerreviewmaketitle

\section{Introduction}

\IEEEPARstart {A}{ccording} to the World Health Organisation, road traffic injury is a top-ten leading cause of death worldwide across all age groups \cite{world2018global}. The Department for Transport (UK) reported $743$ deaths related to car accidents in 2019 \cite{road2017}. Rear-end collisions are mostly attributed to either delayed brake response or lack of braking force due to slower reaction times, when a following driver does not react quickly enough to the behaviour of a lead vehicle, due to inadequate or late detection of its deceleration \cite{winsum1996}. Researchers have examined methods of alerting drivers to avoid rear-end crashes through improved technology either inside or outside the vehicle \cite{Bullough2007,GAO2017,ISLER2010,Li2014,LI2008}.

A majority of traffic safety studies incorporate driver reaction time (RT) in their analysis models of driver behaviour, particularly related to imminent collisions \cite{markkula2016farewell}. 
RT usually represents the time duration measured from the appearance of a stimuli (e.g. a potential hazard such as a lead vehicle’s brake lights activating), until the driver  initiates some form of evasive response \cite{society2015operational} (e.g. depressing the brake pedal). 
Considering braking responses in isolation, effectiveness has traditionally been measured in terms of brake reaction times (BRTs), with common influential factors being driver age, experience, gender, cognitive load and various other stimuli or distractions that a driver needs to consider \cite{barrett1968, jeong2012, schweitzer1995}.

 Effectiveness of various types of stop lamps was also studied \cite{Bullough2000,bullough2001,Sivak1994}, revealing that BRT varies by the lamp type used. 
 Most automotive stop lamp types contain incandescent bulbs, sweeping neon or, increasingly, LED sources. 
 Bullough et al. evaluated these variants for center high-mounted stop lamps (CHMSLs), reporting that incandescent lamps had higher reaction times than LED or neon devices~\cite{Bullough2000}. 
 For incandescent lamps, discernible optical output begins around $50$\,ms after activation, taking around $250$\,ms to reach $90\%$ of steady state output \cite{flannagan1989}. 
 LED CHMSLs also led to shorter RTs since high-luminance point sources naturally provide a stronger stimuli than more diffused sources \cite{Bullough_2001}.

Up to now, as noted, investigations on brake light effectiveness have used brake reaction times (BRT). 
Drivers react differently in various situations; slower at lower speeds, faster in real emergencies, and their responses are affected by issues such as driver height, shoe design, pedal location, seat placement, familiarity, etc. 
To decouple those effects from the influence of the brake design itself; it is necessary to separately measure how quickly a driver perceives the brake light signal, and then how quickly s/he responds.

To explore visual perception factors, vehicle brake lights need to be assessed in terms of their ability to evoke the necessary response or awareness from drivers. 
Eye tracking technology has been employed for this~\cite{hsiao2015}, but suffers from spontaneous responses which do not necessarily involve cognitive perception  (e.g. when the eye glances across a brake light, but the driver is not aware that it has activated).

Brain signals measured by electroencephalogram (EEG), by contrast, are more appropriate measurements, given that any signal has to be firstly recognised by the brain, before a reaction can be made. 
Research on human  perception by Verlerger et al. \cite{verlerger2005} suggested that a specific component in EEG signals called P3 ``reflects a process that mediates between perceptual analysis and response initiation.'' 
In our braking scenario, the `cognitive' component relates to recognition (i.e. awareness of a brake light), which would then be translated into appropriate action (i.e. braking). 
The visual/perceptual component in EEG (known as N1-P1, which occurs earlier around 100\,ms) is related to the perception of a visual stimulus only, so a distraction such as a dazzling light would increase this component but not necessarily result in increased P3 (i.e. not necessarily indicating better recognition of a brake light). 

The present study aims to offer an evaluation of  brake light configurations, varying in shape, size, intensity and type of lamp (incandescent bulb or LED), to determine how brake light design affects elicitation of the most prompt responses. 
This is achieved by extracting and assessing EEG latency component information, with the aim of ensuring a more robust measure of brake light efficacy than BRT.

 As far as we are aware, there have been no extensive studies to date that used \textit{actual} physical brake light assemblies to evaluate the effects of brake light design on the reaction of drivers. 
 Furthermore, this study simulates real driving conditions by  asking subjects to continue depressing the accelerator pedal until they perceive a brake light, at which time they should release the accelerator and depress the brake pedal. 
 This is unlike brake pedal depression timing studies by others which typically employ only a single pedal.
 Our experiments used ten physical brake light assemblies (two pairs with incandescent bulbs and eight pairs containing LEDs, all from recent vehicle models) in a simulation setting, activated with random sequence and timing, using custom built hardware.

\section{Methodology}
\label{sec:method}

\begin{figure}[h!]
\centering
\includegraphics[width=0.6\linewidth]{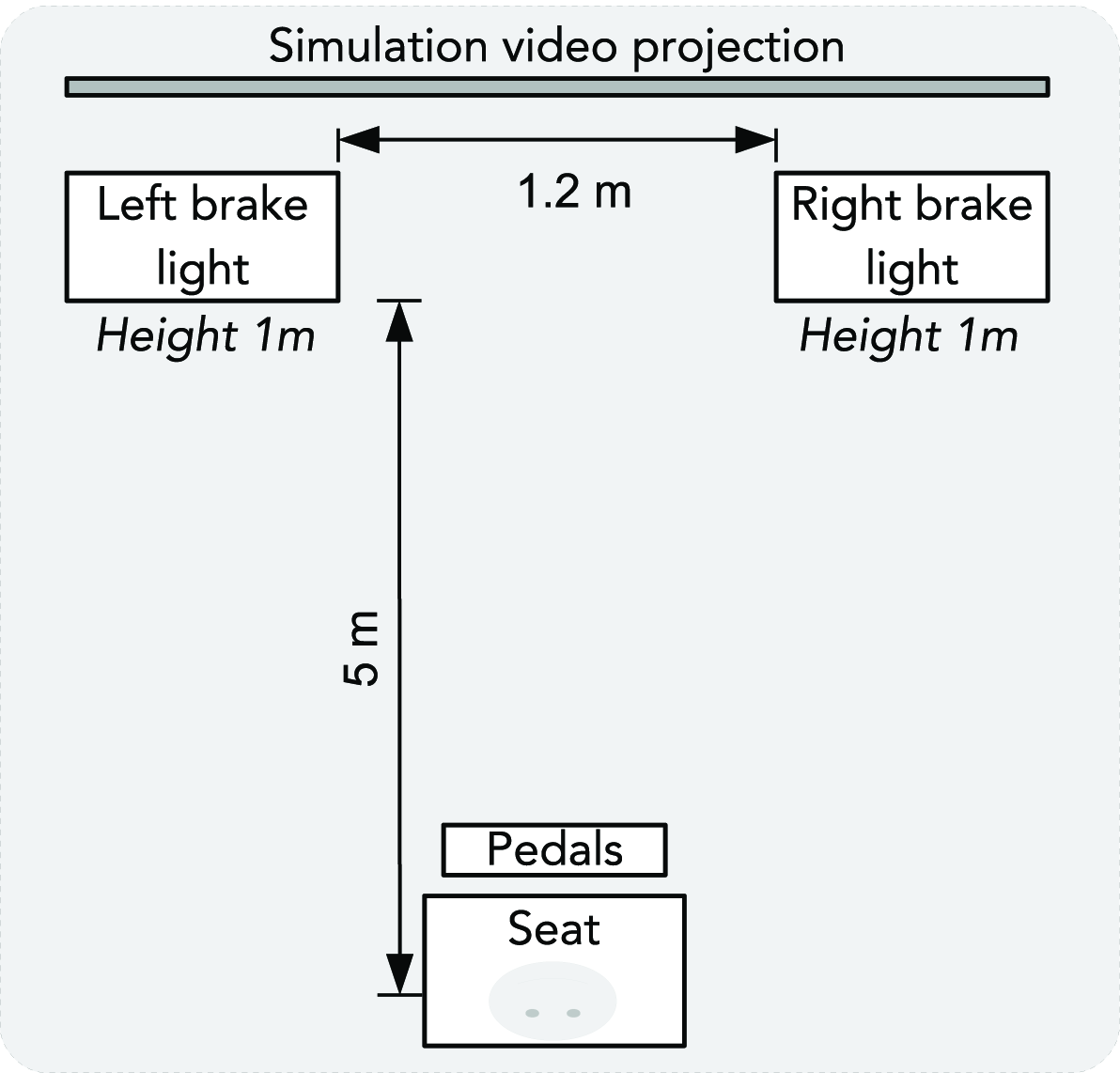}
\caption{Physical layout of the simulation experiments}
\label{layout}
\end{figure}

Experimental hardware and software were constructed to present random brake light events to subjects in a simulated setting, while recording the EEG and responses from a number of associated sensors. 
The experimental data was recorded in a quiet room of size $7.12 \times 14.96$ m  with a projection screen at one end sized $5.00 \times 3.75$ m  for replaying a highway traffic simulation video. Participants for the experiment were seated in an automotive-style chair at a distance of $5$ m  facing the screen as shown in Figure \ref{layout}. Height and gap between brake lights was designed from averaged physical layout of cars.

\vspace*{-7mm}

\subsection{Experimental Hardware}
\begin{figure}[tb]
\centering
\includegraphics[width=0.95\linewidth]{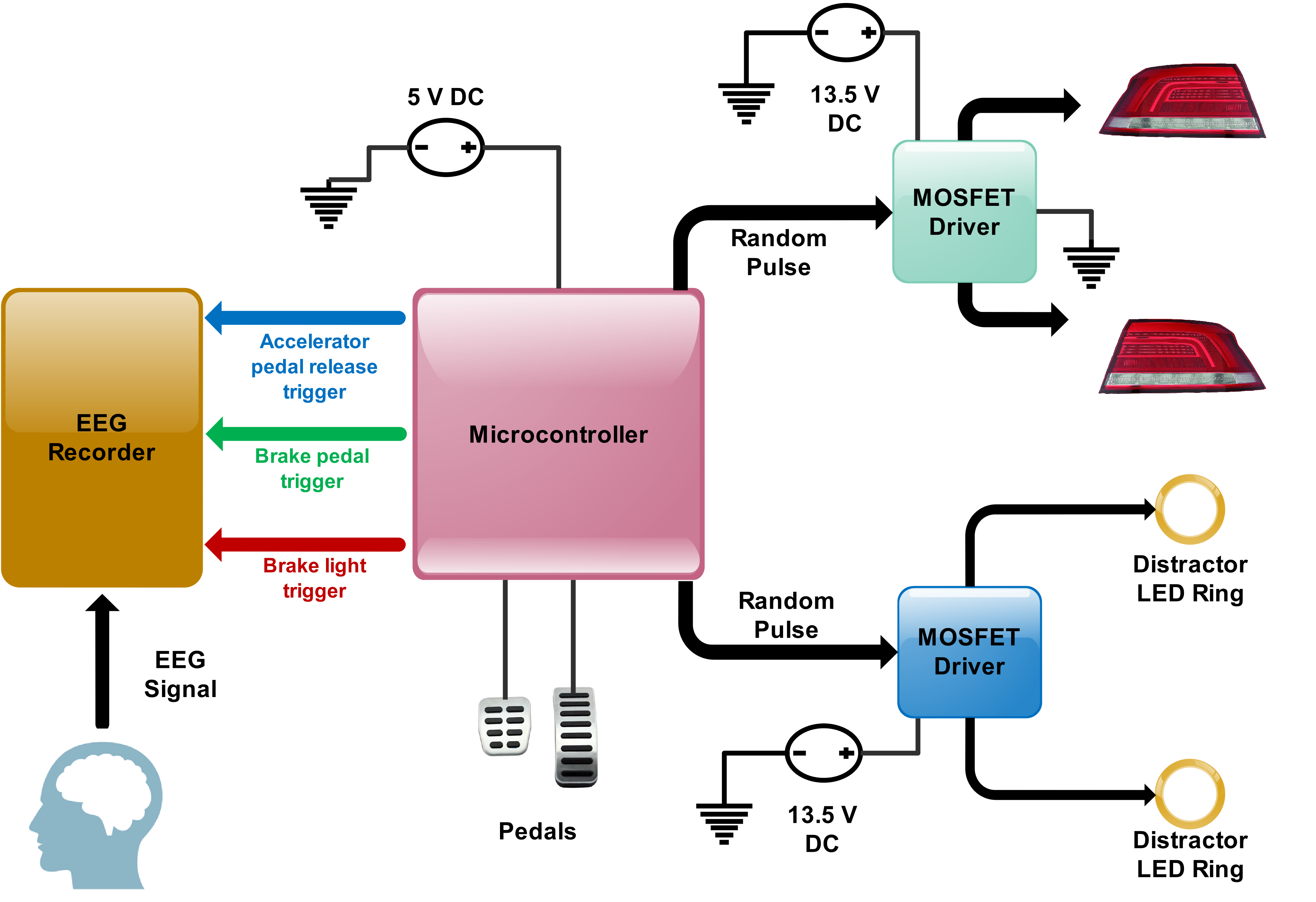}
\caption{Brake light stimulator hardware design blocks.}
 \label{blocks}
\vspace*{-5mm}
\end{figure}

Experimental stimulator hardware was designed using a custom $32$-bit microcontroller system \cite{radial} connected to the pedal switch sensors and two sets of \textsc{MOSFET} driver circuits as shown in Figure \ref{blocks}. 
Maximum turn-on delay for the MOSFET was negligible at only 55 ns. 
One set of \textsc{MOSFETs} was used to drive the brake lights and the other was used for activating two yellow circular rings which were deployed as a distractor. 
The firmware for the system was developed to generating brake light events randomly, while the yellow distractor rings were also illuminated at random intervals, more frequently than the brake lights, but not simultaneously. 

The yellow ring was included to improve the elicitation of the P3 component and to minimise the expectation of the brake light, the yellow rings were introduced to flash randomly when the brake lights were not activated (the firmware was developed to flash only one set of light at any instance - both the brake lights and the distractor yellow rings were activated with random ON times). This also minimised the situation that the participants directly stare at the brake lights waiting for them to light up, which would not occur in reality. With the introduction of random flashes generated by the 100mm diameter yellow rings to distract the attention of the participants, the brake light stimulation introduced more unpredictability, as it would in a real-world situation. 

An event recorder captured all of the timestamped signal information for later analysis.
The collected information consisted of time-stamped brake activation, as well as times from the two foot pedal switches. Further details on the hardware can be found from the project website (https://brake-light.uk).

\vspace*{-3mm}

\subsection{Experimental Setup}

\vspace*{-1mm}

Volunteers were seated in an automotive style char, with an accelerator and brake foot pedal assembly (QLOUNI Industrial Foot-switch Momentary Metal Foot Pedal, part number: $611702431551$),  mounted in front of the seat as shown in Figure \ref{layout}. 
The custom stimulator hardware was programmed to generate $45$ brake light events to turn on (and then off) the brake lights, and similarly to activate the $100$ mm diameter yellow distractor rings in random order.
Brake light activation occurred at random times, and for random periods of between $2$ to $4$ s, with the distractor activation being at random times, illuminated for between $3$ and $5$ s each time, with the constraint that the distractors and brake lights were not activated simultaneously. 
EEG was recorded using OPENBCI hardware kit with eight channels based on the international 10/20 standard at locations F3, Fz, F4, C3, Cz, C4, Pz and Oz, although only channel Pz is used in the analysis here.


\begin{figure}[t]
\centering
\includegraphics[width=3.2in,height=3.2 in,keepaspectratio]{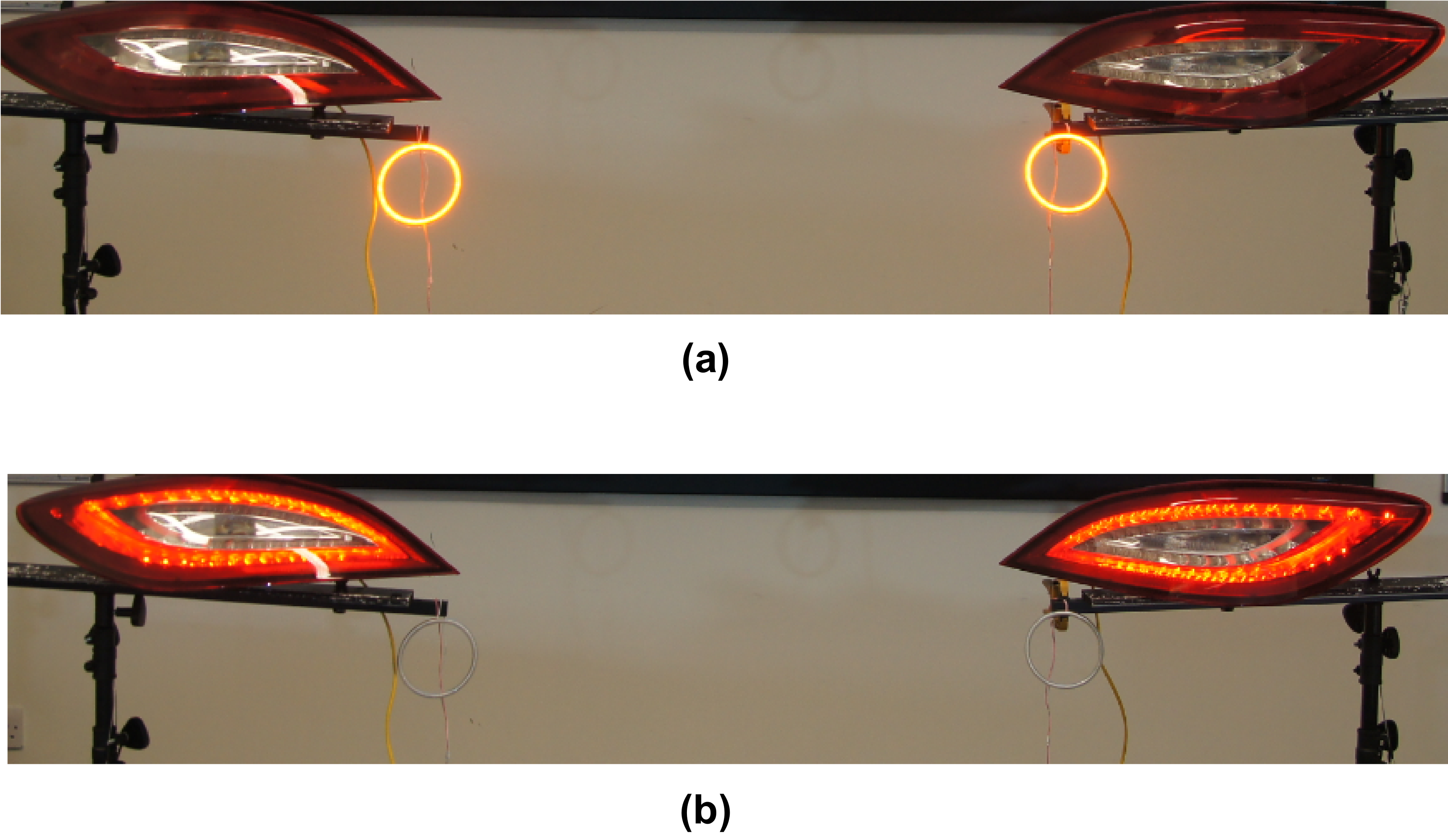}
\caption{Experimental design: (a) Yellow distractor ring with unlit Mercedes brake light (b) Mercedes brake light activated and distractor rings unlit.}
\label{mercedes}
\vspace*{-5mm}
\end{figure}

 Ten sets of physical brake light assemblies from different car manufacturers, selected to represent a range of distinct light shapes from common models, were used in the experiments.  
 Table \ref{table:Table 1} lists the part numbers of the assembly units and bulbs while Figure \ref{mercedes} shows one of the light pairs mounted for this study. Figure 3(a) shows the distractor rings when illuminated, while Figure 3(b) shows the activated brake light.

\begin{table}[h!]
\caption{Details of brake light assemblies used in the experiments}
\label{table:Table 1}
\centering
\begin{tabular}{lccc}
\hline
\textbf{Manufacturer} &
  \textbf{Vehicle} &
  \textbf{Part number} &
  \textbf{Bulb type}\\ \hline
Ford (Bulb) &
  \begin{tabular}[c]{@{}c@{}}Focus\\ (2018)\end{tabular} &
  \begin{tabular}[c]{@{}c@{}}1825320\\ 1825318\end{tabular} &
  Red 1490659 \\ \hline
Fiat (Bulb) &
  \begin{tabular}[c]{@{}c@{}}Fiat 500\\ (2007)\end{tabular} &
  \begin{tabular}[c]{@{}c@{}}OEN 52007424\\ OEN 52007422\end{tabular} &
  \begin{tabular}[c]{@{}c@{}}OSRAMTAIL\\ B001497\end{tabular} \\ \hline
Audi (LED) &
  \begin{tabular}[c]{@{}c@{}}Q5\\ (2016)\end{tabular} &
  \begin{tabular}[c]{@{}c@{}}8R0945093C\\ 8R0945094C\end{tabular} &
  Audi-LED \\ \hline
Fiat (LED) &
  \begin{tabular}[c]{@{}c@{}}Fiat 500\\ (2007)\end{tabular} &
  \begin{tabular}[c]{@{}c@{}}OEN 52007424\\ OEN 52007422\end{tabular} &
  82CRCANR-1 \\ \hline
Ford (LED) &
  \begin{tabular}[c]{@{}c@{}}Focus\\ (2017)\end{tabular} &
  \begin{tabular}[c]{@{}c@{}}1825320\\ 1825318\end{tabular} &
  82CRCANR-1 \\ \hline
Honda (LED) &
  \begin{tabular}[c]{@{}c@{}}Civic\\ (2015)\end{tabular} &
  \begin{tabular}[c]{@{}c@{}}ULT514226\\ ULT514202\end{tabular} &
  PY21W LED \\ \hline
Mercedes (LED) &
  \begin{tabular}[c]{@{}c@{}}CLS-218\\ (2015)\end{tabular} &
  \begin{tabular}[c]{@{}c@{}}OEN A2189067800\\ OEN A2189067700\end{tabular} &
  Benz-LED \\ \hline
Alfa Romeo (LED) &
  \begin{tabular}[c]{@{}c@{}}Mito\\ (2019)\end{tabular} &
  \begin{tabular}[c]{@{}c@{}}LL0604\\ LL0605\end{tabular} &
  LED P21W \\ \hline
Nissan (LED) &
  \begin{tabular}[c]{@{}c@{}}Leaf\\ (2010)\end{tabular} &
  \begin{tabular}[c]{@{}c@{}}OEN 265503NL0A\\ OEN 265553NL0A\end{tabular} &
  Nissan-LED \\ \hline
Volkswagen (LED) &
  \begin{tabular}[c]{@{}c@{}}Golf\\ (2017)\end{tabular} &
  \begin{tabular}[c]{@{}c@{}}5G0945208C\\ 5G0945207C\end{tabular} &
  VW-LED \\ \hline
\end{tabular}
\vspace*{-1mm}
\end{table}

 Eight of the light assemblies contained LED sources, while the remaining two employed incandescent bulbs. 
 In order to make the LED/bulb comparison fairer, we included two same-vehicle model assemblies with different bulb types. Specifically, these were two sets of Ford Focus hatchback and Fiat $500$ units. For both models, we tested one pair of lights that contained incandescent bulbs and another set that used LED sources, but were otherwise identical in size and shape.

\vspace*{-3mm}

\subsection{Experimental Protocol}

The particular brake light unit pair under test were fitted to the mounts, aligned and tested before experimental subject were seated as shown in Figure~\ref{layout}. All experiments were conducted in daylight. A motorway (UK highway) video was projected on the screen, accompanied by the natural traffic and vehicle sounds as recorded -- including tyre, engine and wind noise from the interior of the simulation vehicle as well as from passing vehicles. 
Subjects were given a task during the test, with the aim of keeping their attention focused on the road.  
Specifically, they were asked to keep count of the number of times brake lights were illuminated by other vehicles during the session. 

Each session was designed as a simulated driving paradigm with the brake light assembly in front of the participant representing the \textit{leading vehicle}. Those brake lights were activated at random intervals as noted above. Subjects were instructed to continuously depress the accelerator pedal until they perceived an activation of the brake light in the simulated leading vehicle. 
At that point they were told to immediately release the accelerator and depress the brake pedal. 
They were asked to ignore any flashes or activations of the yellow distractor rings.

The experiment consisted of two sessions, taking place on separate days, each evaluating the efficacy of five different brake light configurations. The sequence in which the lights were presentation to subjects within sessions was randomised.

Data was recorded from a total of $22$ volunteers (age $27.4\pm 5.9$ years, gender balanced). All possessed valid UK driving licenses and had normal or corrected-to-normal vision. 
Half of the subjects were classed as experienced drivers, with more than four years of driving experience. All volunteers were naive to the study and recruited from the local area, and were compensated with £$100$ (£$50$ for each session) in gift vouchers for their time.
 Ethics approval for the protocol was obtained in advance from the University of Kent Faculty of Science Research Ethics committee. We asked subjects to participate in the study only if they were alert and monitored the subjects during experiments and found that all subjects completed every opportunity of braking, i.e. not a single event was missed thereby indicating that the subjects were alert.
 
\vspace*{-3mm}
\subsection{Data Analysis}

 Data analysis was based on brain responses evoked by the different brake light simulations. 
 The brain response component P3 (also known as P300) is evoked during decision making processes that occurs in the brain, in this case when the subjects decided to lift their foot from the accelerator and depress the brake pedal. 
 The P3 component is maximal at mid-line parietal and hence Pz channel was selected~\cite{polich}.

The Pz EEG data was filtered from 0.1 to 8 Hz using an IIR filter to remove the baseline noise, and moreover because P3 is predominantly a low frequency component. 
Next, the EEG was segmented into 45 segments, each corresponding to one brake light activation (since there were 45 brake light activations per brake light for each subject). Each segment of 1.2 seconds was obtained for the period of 0.2 seconds before the brake light onset and 1 second afterwards, which was sufficient to capture the evoked brain responses. 
The segments were ensemble averaged to reduce EEG components that are not time-locked to the brake light cognitive processing. 
This is because the amplitudes of P3 components evoked by the response to the brake lights are very small compared to ongoing brain activity. 
Averaging the 45 EEG segments resulted in amplification of this component (as it is somewhat time locked) and rejection of uncorrelated signals. 
The pre-stimulus baseline of 0.2 s of the EEG pre-onset was used to baseline the post-onset 1 second EEG (i.e. removing the mean using the pre-stimulus baseline). 
A maximal peak around 300 to 600 ms was obtained and the time when this peak occurred (measured from stimulus onset) yielded the response latency. 

With 22 subjects and 10 brake lights, there were 22 x 10 averaged latency values. The outputs of all analysis measures were subjected to Friedman and Kruskal-Wallis tests (with \textit{$\alpha=0.05$} as significance threshold) to gauge statistical significance, since the normality of data distribution was not assumed. 
Post-hoc Wilcoxon signrank testing with Bonferroni corrections were then applied where significant differences in the pre-hoc test was indicated, and thus determine any significant pair-wise differences. 
The overall hypothesis is that efficient brake lights will induce quicker cognitive responses (i.e. lower latencies). 

Figures 4 and 5 show examples of the ensemble averaged brake light signals from one subject for the Ford bulb and LED assembly tests, respectively. The evoked P3 component can be seen as marked, with the latency (time delay) indicated by the double arrows. It quite clear in this instance that the Ford LED lights' P3 latency is lower for that subject than the latency from the Ford bulb assembly.

\begin{figure}[h]
\centering
\includegraphics[width=3.2in,height=3.2in,keepaspectratio]{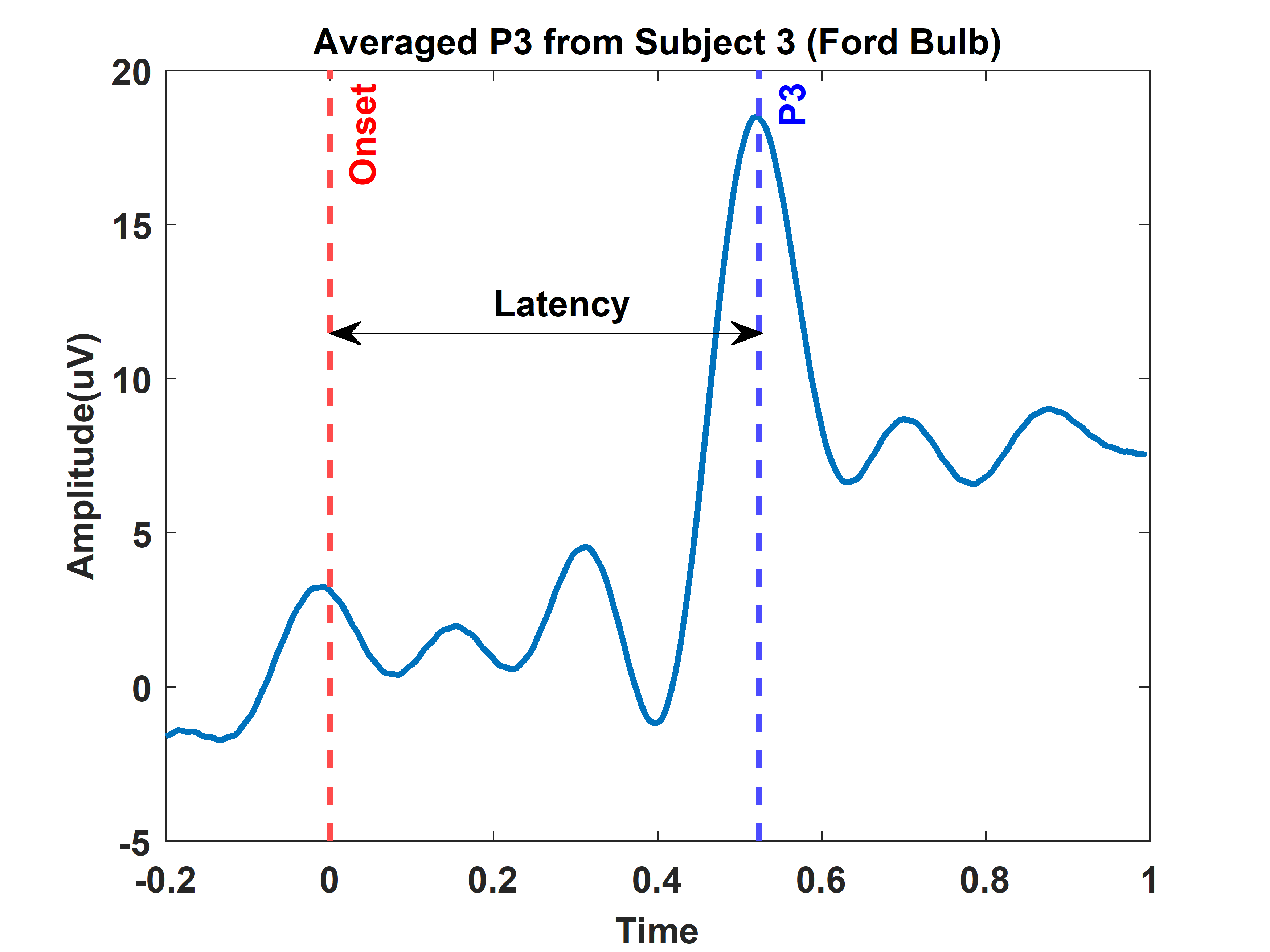}
\caption{Averaged P3 plot from a subject for Ford Bulb.}
\label{P3a}
\vspace*{-5mm}
\end{figure}

\begin{figure}[h]
\centering
\includegraphics[width=3.2in,height=3.2in,keepaspectratio]{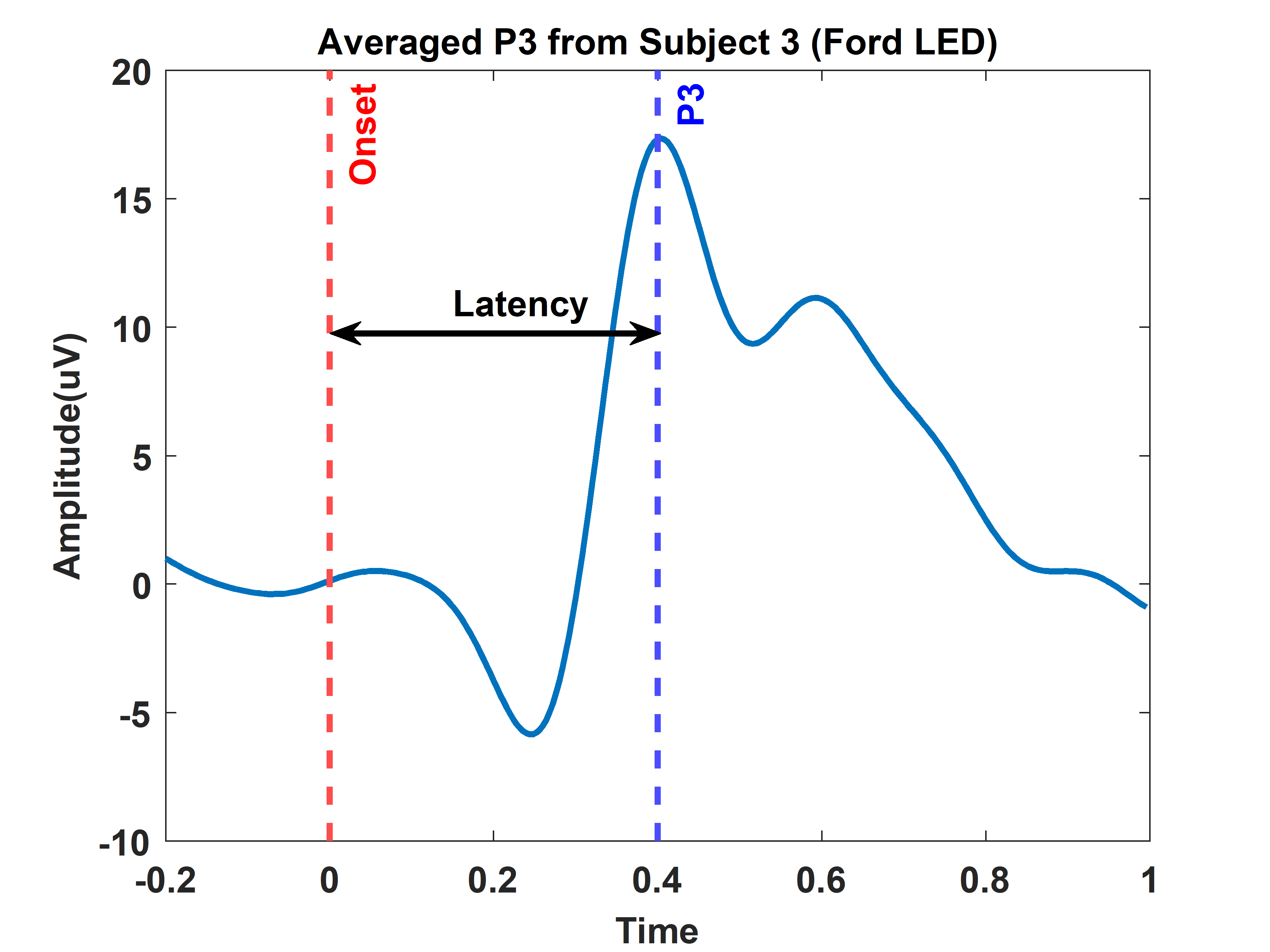}
\caption{Averaged P3 plot from a subject for Ford LED.}
\label{P3b}
\vspace*{-5mm}
\end{figure}

\section{Results and Discussions}
\label{sec:results}

Figure 6 plots the latency of the measured P3 components for all tested brake light assemblies. 
There was statistical difference between the P3 latencies from the different brake lights (Friedman,  ${\chi^2}$(9)=86.1, p=9.91e-15). The bulb-based units were statistically slower than LED-based units (i.e. the average of the two bulbs and average of the eight LED lights; Wilcoxon signrank, Z=4.09, p=2.14e-5).

Of the bulb units, the Fiat bulb had a lower P3 latency than that of the Ford (Wilcoxon signrank, Z=2.06, p=1.96e-1). The latter had the highest P3 latency of all the tested assemblies.

\begin{figure}[h]
\centering
\includegraphics[width=3.2in,height=3.2in,keepaspectratio]{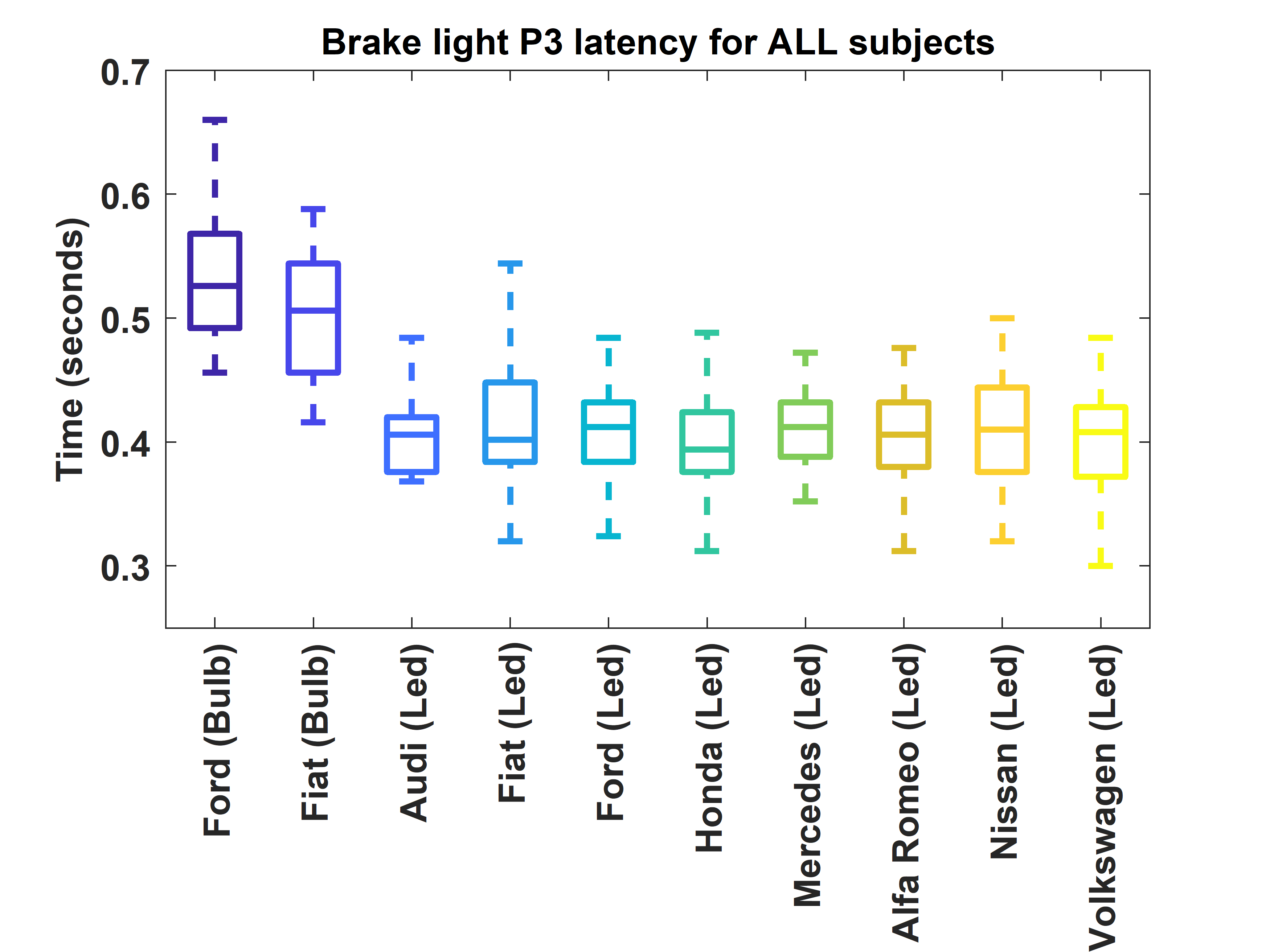}
\caption{Latency of P3 components from EEG for all brake lights.}
\label{P3latency}
\vspace*{-2mm}
\end{figure}

 Within the LED lights, there were differences although not statistically significant (Friedman,  ${\chi^2}$(7)=2.30, p=9.41e-1). We speculate this could be due to noise in the P3 response, where the cognitive component related to the perception and recognition of the brake activation might be confounded with movement related artifact, as subjects were lifting the leg from the accelerator and depressing the brake pedal. Nevertheless, the fact that there is strong statistical difference between the bulb and LED brake lights for the P3 latency does show that the LED based brake lights have a clear advantage in their effectiveness to draw quicker response from subjects.
 
\begin{table*}[!htbp]
\centering
\caption{Mean and standard deviation of P3 latencies for the different brake lights}
\label{table:Table 2}
\resizebox{\textwidth}{!}{%
\begin{tabular}{l | c c | c c c c c c c c}
 &
  \multicolumn{2}{c|}{\textbf{Bulb}} &
  \multicolumn{8}{c}{\textbf{LED}} \\

\textbf{Subjects} & Ford & Fiat & Audi & Fiat & Ford & Honda & Mercedes & Alfa Romeo & Nissan & Volkswagen \\ \hline
Experienced & 
  $0.53\pm0.06$ &
  $0.51\pm0.05$ &
  $0.42\pm0.05$ &
  $0.41\pm0.06$ &
  $0.41\pm0.05$ &
  $0.41\pm0.02$ &
  $0.41\pm0.03$ &
  $0.42\pm0.03$ &
  $0.42\pm0.06$ &
  $0.41\pm0.03$ \\ 
Inexperienced &
  $0.54\pm0.05$ &
  $0.50\pm0.05$ &
  $0.40\pm0.07$ &
  $0.41\pm0.04$ &
  $0.40\pm0.06$ &
  $0.39\pm0.05$ &
  $0.41\pm0.04$ &
  $0.41\pm0.08$ &
  $0.41\pm0.07$ &
  $0.40\pm0.08$ \\\hline
All &
  $0.54\pm0.05$ &
  $0.50\pm0.06$ &
  $0.41\pm0.05$ &
  $0.41\pm0.05$ &
  $0.41\pm0.05$ &
  $0.40\pm0.04$ &
  $0.41\pm0.03$ &
  $0.41\pm0.06$ &
  $0.42\pm0.07$ &
  $0.41\pm0.06$ \\

\end{tabular}%
}
\end{table*}
 
The experience level of subjects did not show any significance with incandescent brake lights (with type of brake light effects removed; Friedman test, ${\chi^2}$(1)=0.124, p=7.25e-1). 
However, when considering the effects of LED based brake lights, there was marginal significance (Friedman,  ${\chi^2}$(1)=3.57, p=5.88e-2). 
We speculate that since the bulb is relatively slow to evoke a  response from both experienced and inexperienced subjects, the difference is not so apparent. 
However, experience does matter somewhat when LED lights are considered -- inexperienced subjects are slower to respond. 
As a further evidence of this, the probability plot of Figure 7 that compares the distribution of the data to the normal distribution. 
This reveals that inexperienced subjects are much slower to respond, shown by the smaller gradient of the normal red line compared to the experienced subjects (for example, if 95\% is considered, experienced subjects have a P3 latency of about 0.5 seconds while inexperienced subjects have a latency of about 0.55 seconds). 

\begin{figure}[h]
\centering
\includegraphics[width=3.2in,height=3.2in,keepaspectratio]{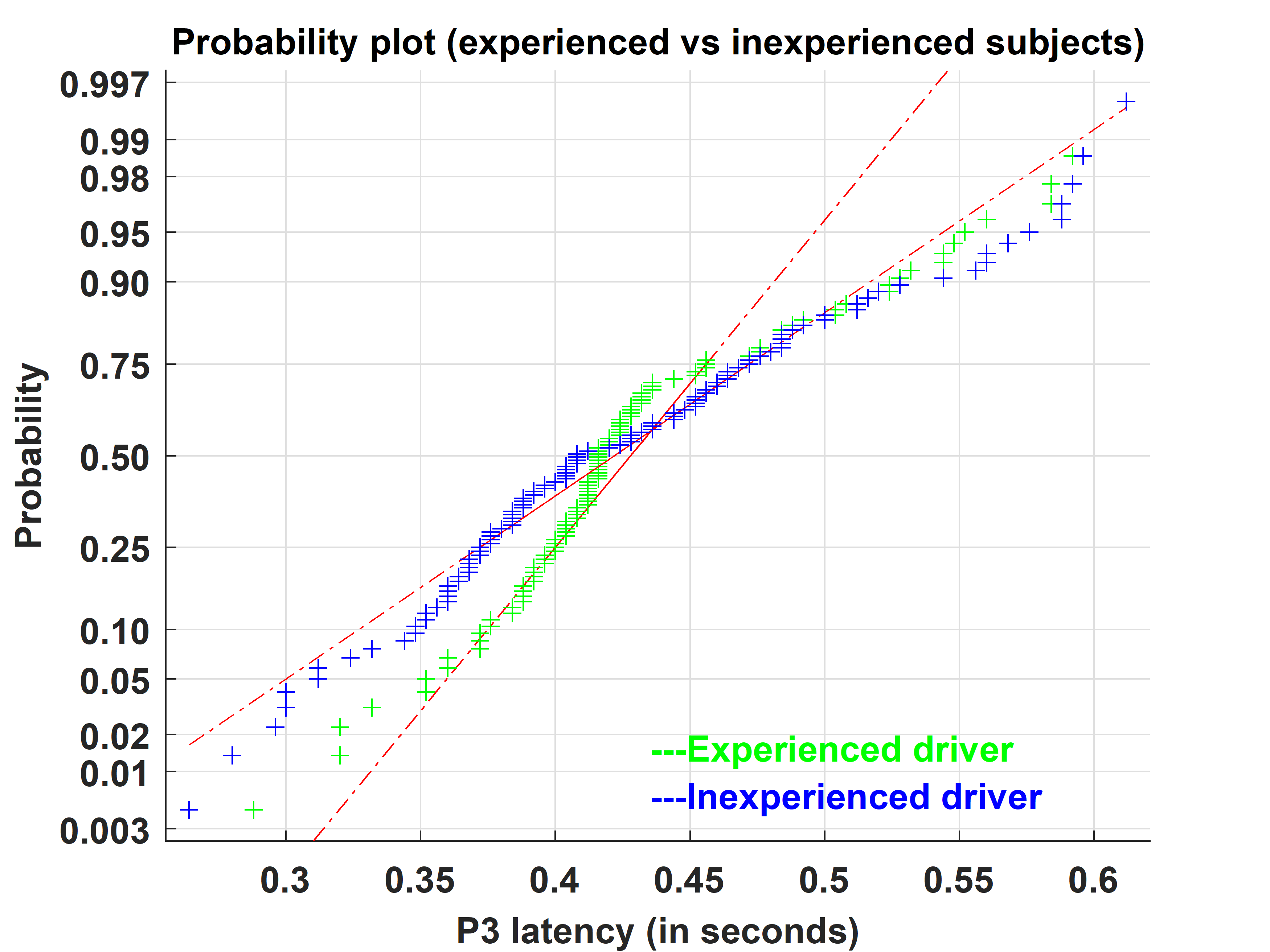}
\caption{P3 latency probability (experienced vs inexperienced) for all brake lights.}
\label{P3 exp}
\vspace*{-3mm}
\end{figure}

Figure 8 plots P3 latency versus the driving experience of all subjects (in months). There is no significant correlation between the P3 latency and experience, which could be due to the artifact issues mentioned earlier (adjusted $r^2$=-0.04, p=0.665).  
Similarly, Figure 9 plots P3 latency versus the age of the subjects in years, where there is no significant correlation as well (adjusted $r^2$=-0.05, p=0.85). The similarity of results from age and experience is as expected given that the age of subjects and their experience were significantly correlated (adjusted r2 =0.36, p=0.0019) - as shown in Figure 10.  

\begin{figure}[h]
\centering
\includegraphics[width=3.2in,height=3.2in,keepaspectratio]{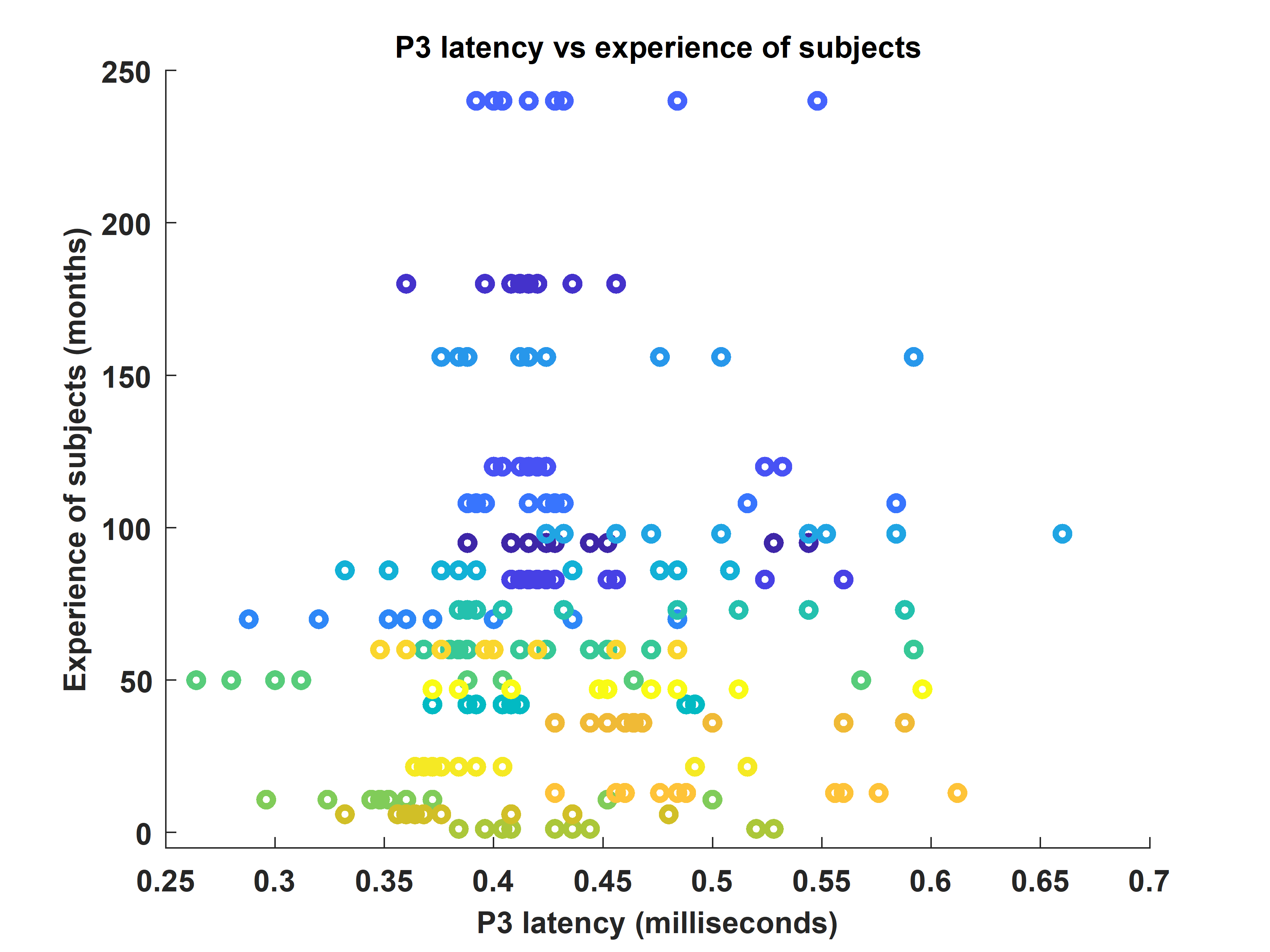}
\caption{P3 latencies versus driving experience of subjects for all brake lights.}
\label{fig8}
\vspace*{-4mm}
\end{figure}
\begin{figure}[h]
\centering
\includegraphics[width=3.2in,height=3.2in,keepaspectratio]{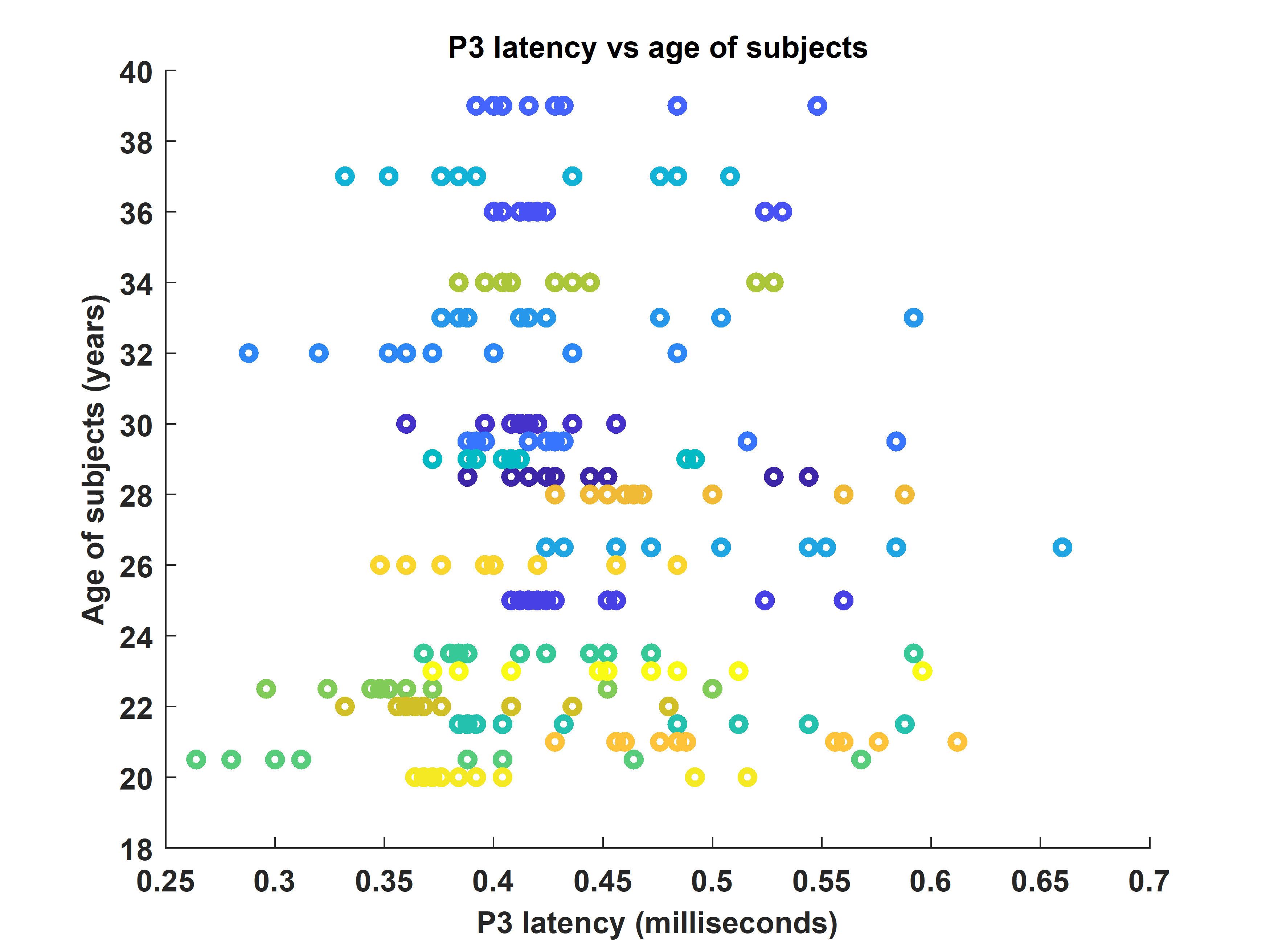}
\caption{P3 latencies versus age of subjects for all brake lights.}
\label{fig9}
\vspace*{-1mm}
\end{figure}

\begin{figure}[h]
\centering
\includegraphics[width=3.2in,height=3.2in,keepaspectratio]{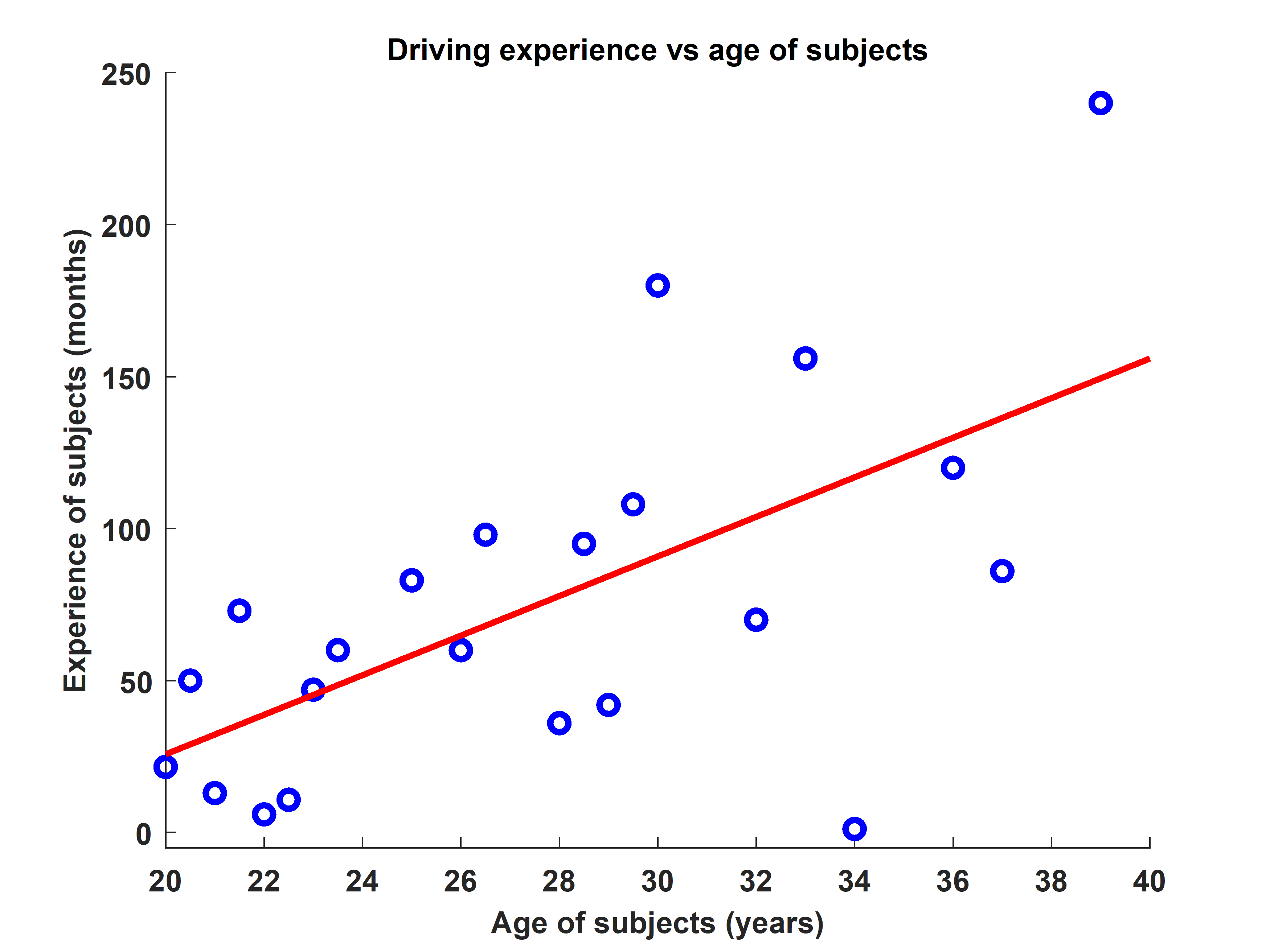}
\caption{Correlation between driving experience vs age of subjects.}
\label{fig9}
\vspace*{-3mm}
\end{figure}

Figure 11 plots all P3 latencies subject-wise, where the first 11 are experienced subjects and the remainder are not experienced. 
Significant differences exist between subjects (Kruskal-Wallis tests, ${\chi^2}$(21)=75.3, 4.87e-8), between experienced subjects (${\chi^2}$(10)=27.4, 2.22e-3) and inexperienced subjects (${\chi^2}$(10)=43.59, 3.91e-6). 
Comparing inexperienced subjects, experienced subjects generally have less difference amongst them as indicated by the higher \textit{p} value in the Kruskal-Wallis statistical tests. 
This is as expected as inexperienced subjects will naturally tend to have larger variance amongst their abilities to respond to the brake lights.

\begin{figure}[h]
\centering
\includegraphics[width=3.2in,height=3.2in,keepaspectratio]{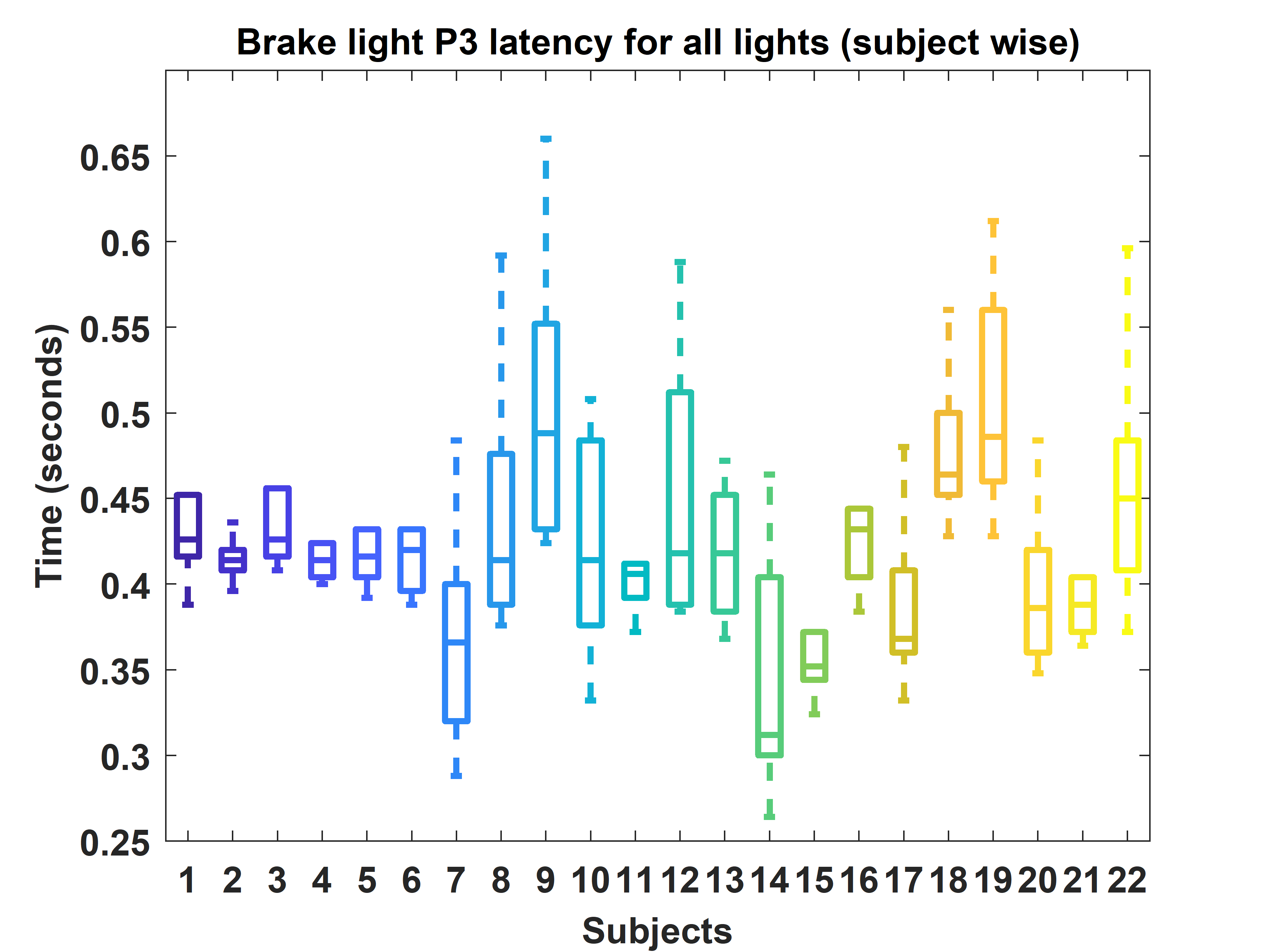}
\caption{P3 latencies for all brake lights (first  11  are  experienced  subjects).}
\label{fig10}
\vspace*{-4mm}
\end{figure}

Table II lists the mean and standard deviation of P3 latencies from all the brake lights. The slowest is the Ford bulb while the fastest (using average P3 latency), is the Honda LED; however this was not statistically significant from other LED lights.  

Although differences between bulb and LED-based brake lights are evident from the results, the study is not without limitations. The use of a video screen to simulate a real driving environment may not accurately represent the driving ability of the subjects. 
Moreover, drivers in a real environment may be more cautious than in the lab environment (as they know there are no consequences of failing to recognise and act on perception of the brake lights). 
However, these are disadvantages common to any laboratory based analysis which simulates a real environment.      

\vspace*{-2mm}

\section{Conclusion}
\label{sec:conclusion}

This study investigated EEG analysis of brake lights based on conventional bulbs and newer LED designs. 
P3 components were analysed from channel Pz for 22 subjects with ten different brake light assemblies, and analysed for statistical differences in terms of the latency of the cognitive component from the brake light onset. 
It was found that both the bulb-based lights evoked slower responses than all of the LED lights, and our recommendation is for bulb-based lights to be replaced by LED counterparts where possible. 
The lack of significant differences in the P3 latency for LED based lights could be attributed to EEG noise caused by movement and other artifacts. 
The results also indicated that experienced subjects were marginally faster than inexperienced subjects, but only when LED lights were considered. 

For our future work, we are planning to develop noise reduction algorithms and to analyse the actual cognitive responses from the braking events using EEG signals in real-life traffic conditions (i.e. live, on the road) as this would allow us to understand the brain processes involved in the recognition of the lights, and the corresponding braking actions. 
Our aim here was to analyse the cognitive response rather than the visual/perceptual component in EEG, hence we limited the analysis to P3 from channel Pz here but in future, we will investigate the interaction between the evoked response components and different activation areas in the brain.  

\vspace*{-2mm}

\section*{Acknowledgment}

We acknowledge the support of the Road Safety Trust (RST 90\_4\_18) in funding this study. The research was conducted with the aim of increasing road safety, and did not have the involvement of any manufacturer. We note that only one type of light assembly from a single vehicle within each manufacturer's range was assessed, and results are therefore not to be interpreted as applying to any other vehicles or brake light assemblies, and certainly not as any endorsement or otherwise of a particular manufacturer. We also acknowledge that the sample sizes here are quite small, so it would be appropriate for replications of the work to confirm the findings. 

\bibliographystyle{IEEEtran}

\bibliography{main}

\begin{thebibliography}{10}
\providecommand{\url}[1]{#1}
\csname url@samestyle\endcsname
\providecommand{\newblock}{\relax}
\providecommand{\bibinfo}[2]{#2}
\providecommand{\BIBentrySTDinterwordspacing}{\spaceskip=0pt\relax}
\providecommand{\BIBentryALTinterwordstretchfactor}{4}
\providecommand{\BIBentryALTinterwordspacing}{\spaceskip=\fontdimen2\font plus
\BIBentryALTinterwordstretchfactor\fontdimen3\font minus \fontdimen4\font\relax}
\providecommand{\BIBforeignlanguage}[2]{{%
\expandafter\ifx\csname l@#1\endcsname\relax
\typeout{** WARNING: IEEEtran.bst: No hyphenation pattern has been}%
\typeout{** loaded for the language `#1'. Using the pattern for}%
\typeout{** the default language instead.}%
\else
\language=\csname l@#1\endcsname
\fi
#2}}
\providecommand{\BIBdecl}{\relax}
\BIBdecl

\bibitem{world2018global}
{World Health Organization}, ``Global status report on road safety 2018: Summary,'' WHO/NMH/NVI/18.20, Tech. Rep., 2018.

\bibitem{road2017}
D.~Robineau, M.~Dark, P.~Baden, A.~Bhagat, A.~Dhani, and H.~Mann, ``Reported road casualities in {G}reat {B}ritain: 2019 complete report,'' Department of Transportation, Tech. Rep., 2019.

\bibitem{winsum1996}
W.~V. Winsum and A.~Heino, ``Choice of time-headway in car-following and the role of time-to-collision information in braking,'' \emph{Ergonomics}, vol.~39, no.~4, pp. 579--592, 1996.

\bibitem{Bullough2007}
J.~D. Bullough, J.~V. Derlofske, and M.~Kleinkes, ``Rear signal lighting: From research to standards, now and in the future,'' in \emph{SAE Paper 2007-01-1229, Warrendale, PA: Society of Automotive Engineers.}\hskip 1em plus 0.5em minus 0.4em\relax SAE International, 2007, pp. 61--69.

\bibitem{GAO2017}
J.~Gao and G.~A. Davis, ``Using naturalistic driving study data to investigate the impact of driver distraction on driver's brake reaction time in freeway rear-end events in car-following situation,'' \emph{Journal of Safety Research}, vol.~63, pp. 195 -- 204, 2017.

\bibitem{ISLER2010}
R.~B. Isler and N.~J. Starkey, ``Evaluation of a sudden brake warning system: Effect on the response time of the following driver,'' \emph{Applied Ergonomics}, vol.~41, no.~4, pp. 569 -- 576, 2010.

\bibitem{Li2014}
G.~Li, W.~Wang, S.~E. Li, B.~Cheng, and P.~Green, ``Effectiveness of flashing brake and hazard systems in avoiding rear-end crashes,'' \emph{Advances in Mechanical Engineering}, vol.~6, p. 792670, 2014.

\bibitem{LI2008}
Z.~Li and P.~Milgram, ``An empirical investigation of a dynamic brake light concept for reduction of rear-end collisions through manipulation of optical looming,'' \emph{International Journal of Human-Computer Studies}, vol.~66, no.~3, pp. 158 -- 172, 2008.

\bibitem{markkula2016farewell}
G.~Markkula, J.~Engstr{\"o}m, J.~Lodin, J.~B{\"a}rgman, and T.~Victor, ``A farewell to brake reaction times? {K}inematics-dependent brake response in naturalistic rear-end emergencies,'' \emph{Accident Analysis \& Prevention}, vol.~95, pp. 209--226, 2016.

\bibitem{society2015operational}
{Society of Automotive Engineers}, ``Operational definitions of driving performance measures and statistics,'' Society of Automotive Engineers, Warrendale, PA, Tech. Rep., 2015.

\bibitem{barrett1968}
G.~V. Barrett, ``Feasibility of studying driver reaction to sudden pedestrian emergencies in an automobile simulator,'' \emph{Human Factors}, vol.~10, no.~1, pp. 19--26, 1968.

\bibitem{jeong2012}
H.~Jeong and P.~Green, ``Forward collision warning modality and content: A summary of human factors studies,'' Technical Report UMTRI-2012-35, Tech. Rep., 2012.

\bibitem{schweitzer1995}
N.~Schweitzer, Y.~Apter, G.~Ben-David, D.~G. Liebermann, and A.~Parush, ``A field study on braking responses during driving. ii. minimum driver braking times,'' \emph{Ergonomics}, vol.~38, no.~9, pp. 1903--1910, 1995.

\bibitem{Bullough2000}
J.~D. Bullough, P.~R. Boyce, A.~Bierman, K.~M. Conway, K.~Huang, C.~P. O’Rourke, C.~M. Hunter, and A.~Nakata, ``Response to simulated traffic signals using light-emitting diode and incandescent sources,'' \emph{Transportation Research Record}, vol. 1724, no.~1, pp. 39--46, 2000.

\bibitem{bullough2001}
J.~Bullough, M.~Rea, R.~Pysar, H.~Nakhla, and D.~Amsler, ``Rear lighting configurations for winter maintenance vehicles,'' in \emph{IESNA Annual Conference: Ottawa, ON, Canada}, 2001, pp. 87--94.

\bibitem{Sivak1994}
M.~Sivak, M.~J. Flannagan, T.~Sato, E.~C. Traube, and M.~Aoki, ``Reaction times to neon, {LED}, and fast incandescent brake lamps.'' \emph{Ergonomics}, vol. 37 6, pp. 989--94, 1994.

\bibitem{flannagan1989}
M.~Flannagan and M.~Sivak, ``An improved braking indicator,'' \emph{SAE Transactions}, pp. 284--288, 1989.

\bibitem{Bullough_2001}
J.~D. Bullough, J.~V. Derlofske, and H.~Yan, ``Evaluation of automotive stop lamps using incandescent and sweeping neon and {LED} light sources,'' in \emph{SAE Paper 2001-01-0301, Warrendale, PA: Society of Automotive Engineers.}\hskip 1em plus 0.5em minus 0.4em\relax SAE International, 2001, pp. 12--19.

\bibitem{hsiao2015}
Y.-C. Hsiao, H.~Kitagawa, and J.~Watada, ``Studies on eye tracking and brainwave measurement,'' in \emph{2015 10th Asian Control Conference (ASCC)}.\hskip 1em plus 0.5em minus 0.4em\relax IEEE, 2015, pp. 1--6.

\bibitem{verlerger2005}
R.~Verleger, P.~Ja{\'s}kowski, and E.~Wascher, ``Evidence for an integrative role of {P3b} in linking reaction to perception,'' \emph{Journal of Psychophysiology}, vol.~19, no.~3, pp. 165--181, 2005.

\bibitem{radial}
S.~{Mouli} and R.~{Palaniappan}, ``Radial photic stimulation for maximal {EEG} response for {BCI} applications,'' in \emph{2016 9th International Conference on Human System Interactions (HSI)}, 2016, pp. 362--367.

\bibitem{polich}
J.~Polich, ``Oxford library of psychology. {T}he {O}xford handbook of event-related potential components,'' {S.J.Luck and E.S.Kappenman}, Ed.\hskip 1em plus 0.5em minus 0.4em\relax Oxford University Press, 2012, ch. Neuropsychology of P300, pp. 159--188.

\end{thebibliography}

\end{document}